\documentclass[aps,prl,twocolumn,showpacs,superscriptaddress]{revtex4}
\usepackage{graphicx}
\usepackage{amsmath}
\usepackage{amssymb}
\usepackage{aas_macros}

\begin{document}

\title{Robust identification of isotropic diffuse gamma rays from Galactic dark matter}

\author{Jennifer M. Siegal-Gaskins}
\affiliation{Center for Cosmology and Astro-Particle Physics, Ohio State University, Columbus, OH 43210}
\author{Vasiliki Pavlidou}
\affiliation{Einstein (GLAST) Fellow}
\affiliation{%
Astronomy Department, California Institute of Technology, Pasadena, CA 91125}%

\date{June 18, 2009}

\begin{abstract}
Dark matter annihilation in Galactic substructure will produce diffuse gamma-ray emission of remarkably constant intensity across the sky, making it difficult to disentangle this Galactic dark matter signal from the extragalactic gamma-ray background.  We show that if Galactic dark matter contributes a modest fraction of the measured emission in an energy range accessible to the Fermi Gamma-ray Space Telescope, the energy dependence of the angular power spectrum of the total measured emission could be used to confidently identify gamma rays from Galactic dark matter substructure.
\end{abstract}

\maketitle

{\bf Introduction.---}
Many independent observations indicate the presence of a substantial non-baryonic dark matter component in the universe, but its nature remains unknown.  Among the favored candidates for cold dark matter are weakly interacting massive particles such as the neutralino in supersymmetric theories and the lightest Kaluza-Klein particle in theories of universal extra dimensions.  In these scenarios dark matter can annihilate, producing standard model particles, and therefore may be detectable indirectly via its annihilation products, such as gamma rays.  The Large Area Telescope aboard the recently launched Fermi Gamma-ray Space Telescope represents an unprecedented leap in gamma-ray detection capabilities in the GeV energy band.  \emph{Fermi} will provide for the first time a detailed all-sky measurement of diffuse emission from $\sim$ 20 GeV up to several hundred GeV, an energy range which is of particular interest for indirect dark matter searches. 

Gamma rays from Galactic dark matter are produced by annihilation in the smooth dark matter halo and in its substructure.  While a few individual subhalos may be detectable as point sources, most will not be resolved by \emph{Fermi} and will instead contribute to the measured diffuse emission \cite{pieri_bertone_branchini_08,kuhlen_diemand_madau_08}.  Numerical simulations have shown that the radial distribution of subhalos is not concentrated toward the Galactic Center, but instead is quite extended relative to the smooth dark matter halo distribution \cite{gao_white_jenkins_etal_04,diemand_kuhlen_madau_07,kuhlen_diemand_madau_07,madau_diemand_kuhlen_08,springel_wang_vogelsberger_etal_08}.  This leads to the important result that the diffuse emission from annihilation in Galactic substructure as observed from the Earth appears  {\em nearly isotropic} on large angular scales \cite{kuhlen_diemand_madau_08,siegal-gaskins_08,springel_white_frenk_etal_08}.  The diffuse flux from substructure typically exceeds that from the smooth halo for angles $\gtrsim$ a few tens of degrees from the Galactic Center \cite{pieri_bertone_branchini_08,kuhlen_diemand_madau_08,springel_white_frenk_etal_08,fornasa_pieri_bertone_etal_09}, so substructure is the dominant contributor to the Galactic dark matter flux over most of the sky.

Since gamma rays from Galactic substructure would appear in \emph{Fermi} data as part of the isotropic diffuse emission, it is essential to identify ways of separating this signal from the extragalactic gamma-ray background (EGRB).  
Although the dark matter energy spectrum is distinct from that of known extragalactic source classes, confident identification of a dark matter signal on spectral grounds alone is unlikely.  At energies as low as tens of GeV, photon-photon interactions with the UV, optical, and IR backgrounds (the extragalactic background light, EBL) lead to a suppression of the measured extragalactic source spectrum which can mimic the exponential cut-off in the dark matter spectrum which occurs at the energy corresponding to the dark matter particle mass $m_{\chi}$.  This problem is especially severe because the shape and characteristic energies of these spectral features are minimally constrained due to uncertainties in the EBL as well as in the properties of extragalactic source populations and dark matter.

Recently, several studies have considered anisotropies in the diffuse gamma-ray background as a means of distinguishing between plausible source classes.  The angular power spectrum of the diffuse emission has been predicted for various extragalactic source classes \cite{ando_komatsu_06,ando_komatsu_narumoto_etal_07,miniati_koushiappas_di-matteo_07,cuoco_brandbyge_hannestad_etal_08,taoso_ando_bertone_etal_08,fornasa_pieri_bertone_etal_09} as well as for Galactic dark matter substructure \cite{siegal-gaskins_08,fornasa_pieri_bertone_etal_09,ando_09}.

In this Letter, we outline a general strategy for identifying contributions from multiple source populations to the diffuse gamma-ray emission 
using the energy dependence of the angular power spectrum, and consider here the specific challenge of extracting a Galactic dark matter signal from the EGRB\@.  Recognizing that (1) diffuse emission from unresolved extragalactic sources has different small-scale anisotropy properties than emission from Galactic dark matter substructure, and (2) their relative contributions to the total isotropic diffuse emission vary with energy, we show that the contribution from Galactic dark matter substructure to the total measured emission is identifiable as a modulation of the total angular power spectrum as a function of energy.

{\bf The intensity energy spectrum.---}
The intensity energy spectrum $I_E$ describes the photon intensity (photons per area per time per solid angle per energy) as a function of energy $E$.   

For dark matter annihilation, we consider the continuum gamma-ray intensity energy spectrum for which we use the analytic approximation for neutralinos given in \cite{bergstrom_edsjo_ullio_01}.  The gamma-ray flux from annihilation is proportional to the dark matter density squared, so the normalization of the dark matter signal strongly depends on assumptions about the mass distribution in halos and subhalos, and is degenerate with the assumed value of the annihilation cross section.  We have normalized the dark matter spectra in our scenarios to illustrate the potential of the anisotropy energy spectrum to identify a Galactic dark matter component while requiring that our models are both consistent with EGRET's measurement of the EGRB \cite{sreek98} and plausible under standard assumptions about subhalo properties.  

For the extragalactic component of the diffuse isotropic emission, we consider unresolved blazars to be the only important contribution, and emphasize that this is the most relevant guaranteed source class since it is expected to dominate both the intensity of the EGRB above $E \gtrsim 1$ GeV and the angular power spectrum of the EGRB at $\ell \gtrsim 100$.  We obtain the intensity energy spectrum of unresolved blazars as in \cite{pavlidou_venters_08}.  Predictions for the amplitude of the blazar contribution to the isotropic diffuse background vary by more than an order of magnitude (e.g. \cite{nt06} and references therein), so we choose the normalization for our examples from within the range allowed by well-motivated predictions.  The shape of the intensity energy spectrum $I_E$ from a population of unresolved blazars with power-law spectra can be parameterized by the mean spectral slope $\alpha_{0}$ and the width of the spectral index distribution $\sigma_{0}$. We choose these parameters
to be consistent with constraints on the spectral index distribution derived by \cite{venters_pavlidou_07} based on EGRET observations.  Our ``reference blazar model'' has parameters $\alpha_0 = 2.35$ and $\sigma_0 = 0.15$.  To approximate the spectral feature due to EBL absorption, we use the analytic expression given in \cite{horiuchi_ando_06} which applies an energy- and redshift-dependent cut-off to the extragalactic source spectrum, and assume for simplicity that all blazars reside at a redshift $z_{0}$.

{\bf The anisotropy energy spectrum.---}
We define the anisotropy energy spectrum as the value of the angular power spectrum at a fixed multipole $\ell$ as a function of energy; we fix $\ell=100$ in the example scenarios presented.  We consider the angular power spectrum $C_{\ell}$ of intensity fluctuations $\delta I (\psi) \equiv (I(\psi) - \langle I \rangle)/\langle I \rangle$, where $I(\psi)$ is the intensity in the direction $\psi$.
The angular power spectrum is given by $C_{\ell} \! = \! \langle\, | a_{\ell m} |^{2} \rangle$, where $a_{\ell m}$ are determined by expanding $\delta I (\psi)$ in spherical harmonics, $\delta I (\psi)  \! = \!  \sum_{\ell,m} a_{\ell m} Y_{\ell m}(\psi)$.

For Galactic dark matter substructure we use the angular power spectrum for an anti-biased subhalo distribution with mass function slope $\alpha=0.9$ and $M_{\rm min}=10$ M$_{\odot}$ calculated using the procedure of \cite{siegal-gaskins_08}.  We estimate the value of the angular power spectrum of emission from unresolved blazars at $\ell=100$ from Fig.~4 of \cite{ando_komatsu_narumoto_etal_07}.  In general, the predicted amplitude of the angular power spectrum from substructure is much greater than that from blazars.  This can be understood by noting that both source populations consist primarily of point sources, which results in a noise-like angular power spectrum, the amplitude of which is inversely proportional to the number density of sources per solid angle.  As a cosmological source class, blazars have a much higher number density than Galactic substructure, and thus produce less angular power.

The total angular power spectrum from extragalactic sources $C_{\ell}^{\rm EG}$ and Galactic dark matter substructure $C_{\ell}^{\rm DM}$ is 
\begin{equation}
\label{eq:clsum}
C_{\ell}^{\rm tot}=f_{\rm EG}^{2}C_{\ell}^{\rm EG} + f_{\rm DM} ^{2}C_{\ell}^{\rm DM} + 2f_{\rm EG}f_{\rm DM}C_{\ell}^{{\rm EG}\times{\rm DM}},
\end{equation}
where $f_{\rm EG}$ and $f_{\rm DM}$ are the (energy-dependent) fractions of the total emission from extragalactic sources and Galactic dark matter substructure, respectively.  Since extragalactic sources and Galactic dark matter substructure are uncorrelated, the cross-correlation term $C_{\ell}^{{\rm EG}\times{\rm DM}}=0$.  
The $1-\sigma$ statistical uncertainty in the measured angular power spectrum is given by 
\begin{equation}
\label{eq:deltacl}
\delta C_{\ell}^{\rm s} = \sqrt{\frac{2}{(2\ell + 1)\,\Delta\ell\, f_{\rm sky}}} \left(C_{\ell}^{\rm s} + \frac{C_{\rm N}}{W_{\ell}^{2}}\right),
\end{equation}
where $C_{\ell}^{\rm s}$ is the angular power spectrum of the signal (here, $C_{\ell}^{\rm tot}$) and $W_{\ell}  \! = \!  \exp(-\ell^{2}\sigma_{\rm b}^{2}/2)$ is the window function of a Gaussian beam of width $\sigma_{\rm b}$.  We take the fraction of the sky observed $f_{\rm sky}=0.75$, and multipole bins of $\Delta\ell = 100$.  The noise power spectrum $C_{\rm N}$ is the sum of the Poisson noise of the signal and the background, $C_{\rm N} = (4 \pi f_{\rm sky}/N_{\rm s})(1 + (N_{\rm b}/N_{\rm s}))$, where $N_{\rm s}$ and $N_{\rm b}$ are the number of signal and background photons observed.  In this analysis, the EGRB and gamma rays from Galactic dark matter substructure together constitute the signal, while any other observed gamma rays, such as those from Galactic sources other than dark matter substructure, are considered background photons.  To assess the impact of plausible backgrounds on the measurement errors, we assume $N_{\rm b}/N_{\rm s}=10$.  This ratio is a rough but generous estimate of the factor by which the Galactic diffuse emission predicted by the GALPROP conventional model exceeds our isotropic diffuse intensity for latitudes $|b|>20^{\circ}$ for energies up to $\sim$ 50 GeV \cite{strong_moskalenko_reimer_04gal}.  We take the field of view of \emph{Fermi} to be 2.4 sr, and assume an all-sky observation time of 5 years.  The effective area $A_{\rm eff}$ and angular resolution $\sigma_{\rm b}$ of \emph{Fermi} vary over the energy range considered here, so we approximate the energy dependence of these parameters from the estimated \emph{Fermi} performance plots  \footnote{http://www-glast.slac.stanford.edu/software/IS/ glast\_lat\_performance.htm, accessed January 10, 2009}.  

\begin{figure}
\includegraphics[width=.47\textwidth]{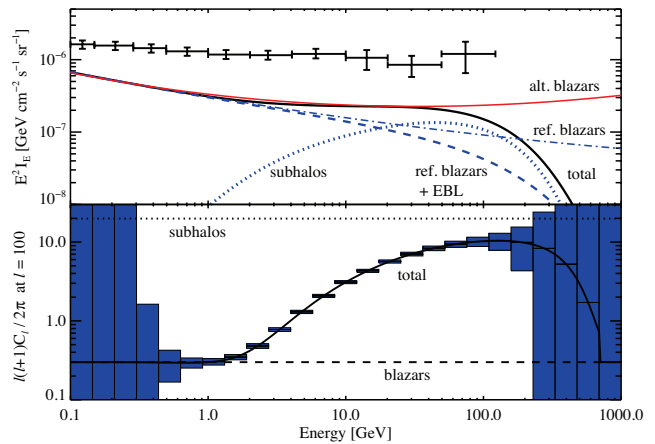}
\caption{\label{fig:ex700gev} {\it Top panel:}  Example measured isotropic diffuse intensity spectrum. Shown individually are the spectra of Galactic dark matter substructure for $m_{\chi}=700$ GeV, the reference blazar model without and with EBL attenuation ($z_0 = 0.4$), and the unattenuated alternative blazar model. The `total' signal is the sum of the attenuated reference blazar spectrum and the dark matter spectrum.  The EGRET measurement of the EGRB is plotted for reference ({\it black crosses}).  {\it Bottom panel:}  
Energy dependence of the angular power spectrum 
of the total isotropic emission at multipole $\ell=100$ for the scenario shown in the top panel.  The anisotropy energy spectrum of Galactic dark matter substructure, unresolved blazars, and the total signal are shown.}  
\end{figure}

Fig.~\ref{fig:ex700gev} shows an example intensity energy spectrum and corresponding anisotropy energy spectrum for $m_{\chi}=700$ GeV, a mass motivated by attempts to explain the recent results of PAMELA and ATIC with dark matter annihilation \cite{boezio_bonvicini_jerse_etal_08,chang_adams_ahn_etal_08}.  The observed intensity energy spectrum is the sum of the EBL-attenuated reference blazar spectrum and the dark matter spectrum, and in this example dark matter dominates the intensity energy spectrum above $\sim$ 20 GeV.  The observed intensity spectrum is, however, also consistent with a blazar-only spectrum with a broader spectral index distribution (an ``alternative blazar model'', $\alpha_{0}=2.32$, $\sigma_{0}=0.26$) that has suffered EBL attenuation.  In light of uncertainties in the properties of blazars, the EBL, and dark matter, the intensity energy spectrum alone is not sufficient to distinguish between these two possibilities.  In this case the anisotropy energy spectrum can break the degeneracy: if unresolved blazars were the sole source of the isotropic diffuse emission, the anisotropy energy spectrum would be constant in energy, but the presence of a dark matter contribution that varies with energy results in a modulation of the anisotropy energy spectrum. 

\begin{figure}
\includegraphics[width=.47\textwidth]{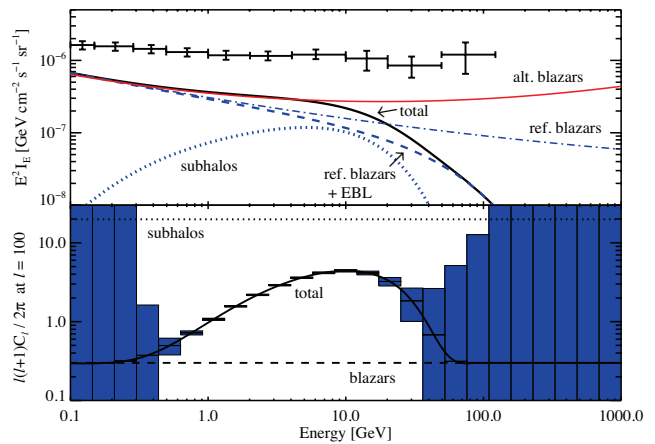}
\caption{\label{fig:ex80gev}  Same as Fig.~\ref{fig:ex700gev}, for $m_{\chi}=80$ GeV.  For the attenuated reference blazar model, $z_{0}=1$.}
\end{figure}

Fig.~\ref{fig:ex80gev} presents a scenario with $m_{\chi}=80$ GeV, which is generally considered a more favorable mass for detection by \emph{Fermi}.  However, in this scenario the dark matter intensity is always subdominant, and as before the observed cut-off in the intensity energy spectrum occurs at an energy consistent with EBL suppression of the EGRB, producing an acute degeneracy between the reference blazar model plus a dark matter contribution and an alternative blazar model ($\alpha_{0}=2.28$, $\sigma_{0}=0.26$) without dark matter.  Again, the anisotropy energy spectrum provides a means of robustly identifying a dark matter contribution: even though Galactic dark matter substructure never dominates the intensity energy spectrum, it produces a strong feature in the anisotropy energy spectrum.

In both examples, the error bars become prohibitively large for $E \lesssim 1$ GeV due to the angular resolution of \emph{Fermi} below this energy, and at sufficiently high energies due to lack of photons.  In between these two regimes, the noise term in Eq.~\ref{eq:deltacl} ($C_{\rm N}/W_{\ell}^{2}$) is negligible, and the uncertainties are quite small, governed primarily by the sample variance at the selected multipole.  As a result, the departure of the measured anisotropy energy spectrum from an energy-invariant quantity can be identified with high confidence, clearly indicating a transition in energy between source populations.

We comment that the blazar intensity spectra (as well as the total intensity) in our examples fall noticeably below the EGRET data points.  This reflects the expectation that \emph{Fermi}, with its enhanced point-source sensitivity, will resolve a large number of extragalactic sources that had contributed to EGRET's measurement of the EGRB, and consequently will measure a lower amplitude diffuse background.  The EGRET data points are plotted to explicitly demonstrate that our models do not violate existing constraints.

{\bf Discussion.---}
The observation of a modulation in the anisotropy energy spectrum robustly indicates a change with energy in the spatial distribution of contributing source population(s).  Although we have considered only the contributions of Galactic dark matter substructure and unresolved blazars to the isotropic diffuse background, sources other than those explicitly considered here (e.g., \emph{Fermi} irreducible backgrounds, the smooth dark matter halo, and additional extragalactic populations including dark matter) which could induce an energy-dependence in the total angular power spectrum are not expected to provide significant power at the angular scales of interest.  Here we have not explicitly considered the possibility of energy dependence of the angular power spectrum of a single source class, although such a dependence could be produced by redshifting of hard features (e.g., lines, exponential cut-offs) in cosmological source spectra \cite{zhang_Beacom_04, ando_komatsu_06}, spatial dependence or redshift evolution of the properties of source populations, or EBL attenuation.  These mechanisms affect the anisotropy energy spectrum by changing the relative contribution of different members of a given source class as a function of energy.  For example, EBL attenuation of high-redshift blazars could reduce the number density of contributing sources at sufficiently high energies.  However, all of these mechanisms for generating strong features in the anisotropy energy spectrum would be accompanied by features (e.g., suppression, in the case of EBL attenuation) in the intensity energy spectrum {\em at the same energies}, so these scenarios can be verified or excluded by joint examination of the intensity and anisotropy energy spectra.  In general, confident identification of a specific source population will require information from the intensity energy spectrum since the modulation must have an energy dependence consistent with expectations for that source population and with features in the intensity energy spectrum.  Moreover, although the noise-like shape of the angular power spectrum from substructure is not unique to this source class, 
the {\it combination} of its shape and amplitude, in addition to the energy spectrum and large-scale angular distribution of substructure (which lacks a strong correlation with Galactic structures), is not reproduced by any other known or predicted source class.

A non-negligible contribution to the isotropic diffuse emission from Galactic dark matter substructure is {\em guaranteed} to produce a modulation of the anisotropy energy spectrum.  The precise shape, amplitude, and prominence of such a feature are governed by the angular power spectra of the EGRB and Galactic dark matter and their contributions to the total intensity, which are subject to considerable uncertainty and thus may differ significantly from the values adopted here, and therefore the scenarios presented should be regarded as examples, not predictions.  However, we emphasize that \emph{any} modulation in the anisotropy energy spectrum can be used to extract information about contributing source populations: a firm prediction for such a feature is not required to infer the presence of multiple or changing populations.

It has not escaped our notice that in the limit where the baseline extragalactic anisotropy level is observable, Eq.~\ref{eq:clsum} provides a novel way of extracting the shape of the dark matter annihilation spectrum even in cases where the dark matter signal cannot be disentangled from other contributions in the observed intensity spectrum.

We are extremely grateful to John Beacom and Stephan Meyer for valuable comments and suggestions, and we thank Shin'ichiro Ando, Stefano Profumo, Luis Reyes, and Tonia Venters for helpful discussions.  We also thank the organizers and participants of the ``Novel Searches for Dark Matter'' workshop hosted by CCAPP at The Ohio State University for stimulating interactions that inspired this work.  VP acknowledges support by NASA through the GLAST Fellowship Program, NASA Cooperative Agreement: NNG06DO90A.

\bibliography{dmiso_v3}

\end{document}